\documentclass[aps,pra,twocolumn,showpacs,superscriptaddress,floatfix]{revtex4}
\usepackage{graphicx}
\usepackage{nicefrac}
\usepackage{amsmath}
\usepackage{amsfonts}
\usepackage{amssymb}
\usepackage{amsthm}
\usepackage{epsf}
\usepackage{bm}
\usepackage{bbm}
\usepackage{longtable}

\usepackage{dcolumn}

\newcolumntype{.}{D{x}{}{-1}}

\newcommand{\balpha}{\vec{\alpha}}

\newcommand{\bfr}{\vec{r}}

\newcommand{\Za}{{Z\alpha}}

\newcommand{\vare}{\varepsilon}

\newcommand{\lbr}{\langle} \newcommand{\rbr}{\rangle}

\begin{document}

\title{QED theory of the nuclear magnetic shielding in hydrogen-like ions}

\author{V.~A. Yerokhin} 
\affiliation{Max~Planck~Institute for Nuclear Physics, Saupfercheckweg~1, D~69117 Heidelberg, Germany}
\affiliation{St.~Petersburg State Polytechnical University, Polytekhnicheskaya 29,
        St.~Petersburg 195251, Russia}

\author{K. Pachucki} \affiliation{Faculty of 
        Physics, University of Warsaw, Ho\.{z}a 69, 00--681 Warsaw, Poland}

\author{Z. Harman} 
\affiliation{Max~Planck~Institute for Nuclear Physics, Saupfercheckweg~1, D~69117 Heidelberg, Germany}

\affiliation{ExtreMe Matter Institute
EMMI, GSI Helmholtzzentrum f\"ur Schwerionenforschung, D-64291 Darmstadt,
Germany}

\author{C.~H. Keitel} 
\affiliation{Max~Planck~Institute for Nuclear Physics, Saupfercheckweg~1, D~69117 Heidelberg, Germany}

\begin{abstract}
The shielding of the nuclear magnetic moment by the bound electron in hydrogen-like ions 
is calculated {\em ab initio} with inclusion of relativistic, nuclear, and
quantum electrodynamics (QED) effects. The QED correction is
evaluated to all orders in the nuclear binding strength parameter and, independently,
to the first order in the expansion in this parameter. The results obtained
lay the basis for the high-precision determination of nuclear magnetic dipole 
moments from measurements of the $g$-factor of hydrogen-like ions. 
\end{abstract}


\maketitle

\begin{figure*}
\centerline{\includegraphics[width=0.8\textwidth]{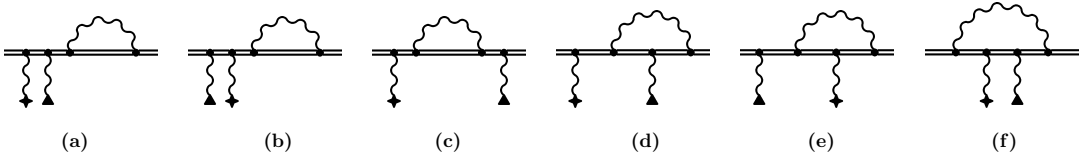}}
\caption{Self-energy correction to the nuclear magnetic shielding. Double line
  represents the electron in the binding nuclear field. Wave line terminated
  by a triangle represents the dipole hyperfine interaction with the nucleus and
  wave line terminated by a cross represents the interaction with the
  external magnetic field.
 \label{fig:1} }
\end{figure*}

Magnetic dipole moments of nuclei are most often determined by the nuclear
magnetic resonance (NMR) technique. Other methods such as 
atomic beam magnetic resonance, collinear laser spectroscopy, and optical
pumping (OP) have also been
used. The measured quantities are usually the ratio of the frequencies
(or the $g$-factors) for the nucleus of interest and the reference
nucleus. Such ratios can be experimentally determined with a part-per-billion (ppb) accuracy
\cite{wineland:83}. However,
magnetic moments of {\em bare} nuclei extracted from these experiments are 
much less accurate. This is because the experimental data should be
corrected for several physical effects, which are difficult to calculate.
The main effect is the diamagnetic shielding of the external magnetic field by
the electrons in the atom. The NMR results should be also corrected
for the paramagnetic chemical shift caused by the chemical environment
\cite{ramsey:50:dia} and the OP data are sensitive to the hyperfine mixing of
the energy levels \cite{lahaye:70}. Significant (and generally unknown)
uncertainties of calculations of these effects often lead to ambiguities in
the published values of nuclear magnetic moments
\cite{gustavsson:98:pra}.

Since the accuracy of calculations of the chemical shifts
cannot be reliably assessed, the means of comparison of nuclear
moments shielded by different environments in NMR measurements 
are rather limited. Independent
determinations of nuclear magnetic moments would
define uncertainties of theoretical calculations of the chemical
shifts and help to assess the accuracy of NMR standards. 

Reliable determination of the nuclear magnetic moments is also prompted by 
a new generation of QED calculations of the hyperfine splitting
in highly charged ions. It was demonstrated \cite{shabaev:01:hfs} that
the magnetic sector of bound-state QED can be tested in these systems
to all orders in the binding field, if the nuclear magnetic moments
are accurately known. Alternatively, comparing theoretical
predictions with experimental results, one can determine nuclear properties and
set benchmark tests for nuclear-structure theory. A recent example is the
spectroscopic determination of the nuclear charge radii of the
neutron-halo nuclei $^8$He, $^{11}$Li, and $^{11}$Be \cite{halo}, which
yielded unique information about the properties of these extraordinary systems.

A way to a high-precision determination of nuclear magnetic moments is to study
the simplest atomic systems, the hydrogen-like ions. Measurements of the
bound-electron $g$-factor in these systems progressed dramatically during the recent
years and reached the ppb level \cite{haeffner:00:prl}. They
led not only to a stringent test of sophisticated QED
calculations \cite{yerokhin:02:prl,pachucki:04:prl} but also to an improved
determination of the electron mass \cite{mohr:05:rmp}. Extensions of these
experiments to ions with a nonzero nuclear spin will provide
a determination of the nuclear magnetic moments from a simple system that can be described
theoretically up to a very high accuracy.

It is well known \cite{bethesalpeter} that the nuclear-spin-dependent
part of the atomic $g$-factor $g_F$ is suppressed by about 3 orders of
magnitude as compared to the leading effect due to the bound-electron 
$g$-factor (see Eq.~(\ref{eq1}) below). This imposes limitations on 
possible determinations of the nuclear magnetic moment from $g_F$.
We show here, however, that the leading
effect cancels exactly in the sum of the $g$-factors for two
hyperfine-structure levels  (see
Eq.~(\ref{eq2}) below). This sum is proportional to the nuclear
$g$-factor and, therefore, is much better suited for extracting
the nuclear magnetic moment.
Its calculation can be conveniently parameterized 
in terms of the nuclear shielding constant $\sigma$, as given by Eq.~(\ref{eq2}).

In this work we perform an {\it ab initio} calculation of
the nuclear magnetic shielding for the ground state of
hydrogen-like ions. The relativistic, QED, and nuclear effects are 
accounted for. The main challenge is the calculation of the QED correction.
To the best of our knowledge, the only attempt to address it
was the estimate reported in Ref.~\cite{rudzinski:09}. In this Letter, we
calculate the QED correction rigorously to all orders in the binding nuclear 
strength parameter $\Za$ (where $Z$ is the nuclear charge and $\alpha$ is the 
fine-structure constant) and, independently, we derive the leading term of
its $\Za$ expansion. 

We now turn to the theory of the $g$-factor of a hydrogen-like ion with a
nonzero spin. Within relativistic quantum mechanics, it is given by 
\cite{bethesalpeter},
\begin{eqnarray} \label{eq1}
g^{(0)}_F = g_j\, \frac{\lbr \bm{j}\cdot\bm{F} \rbr}{F(F+1)}
 - \frac{m}{m_p}\,g_I\, \frac{\lbr \bm{I}\cdot\bm{F} \rbr}{F(F+1)}\,,
\end{eqnarray}
where $F$ is the total angular momentum, $I$ is the nuclear spin, $j$ is the electron
angular momentum, 
$g_j$ is the Dirac bound-electron $g$-factor, $g_I = \mu/(\mu_NI)$ is the nuclear $g$-factor, 
$\mu$ is the nuclear magnetic moment,  
$\mu_N = |e|/(2m_p)$ is the nuclear magneton,
$m$ and $m_p$ are the electron and proton masses, respectively,
$\lbr \bm{j}\cdot\bm{F} \rbr = [F(F+1)-I(I+1)+j(j+1)]/2\,$, and
$\lbr \bm{I}\cdot\bm{F} \rbr = [F(F+1)+I(I+1)-j(j+1)]/2\,$.
The higher-order corrections enter into Eq.~(\ref{eq1}) in two
ways: (i) the Dirac electron $g$-factor $g_j$ is modified by QED and recoil 
effects that do not depend on nuclear spin, 
(ii) the free-nucleus $g$-factor $g_I$  is shielded by the bound electron. 
Additional corrections, e.g., those due to the electric quadrupole interaction
\cite{moskovkin:04}, are small and can be absorbed into the definition of the
nuclear shielding.

For the ground state of an ion with a nuclear spin $I>\nicefrac12$, 
we introduce the combination of $g$-factors $\overline{g}$, 
\begin{eqnarray} \label{eq2}
\overline{g} \equiv g_{F = I+\nicefrac12}+ g_{F = I-\nicefrac12} = -2\frac{m}{m_p}\frac{\mu}{\mu_NI}\,(1-\sigma)\,,
\end{eqnarray}
whith $\sigma$ being the shielding constant. If both $g$-factors $g_{F = I+\nicefrac12}$
and $g_{F = I-\nicefrac12}$ are measured and $\sigma$ is known from theory, the
above formula determines the nuclear magnetic moment $\mu$.
For the ions with a nuclear spin $I=\nicefrac12$, Eq.~(\ref{eq2}) is not applicable
and the nuclear magnetic moment has to be determined from Eq.~(\ref{eq1}).

The nuclear shielding constant $\sigma$ defined by Eq.~(\ref{eq2}) can be
represented as a sum 
\begin{eqnarray} \label{eq3}
\sigma = \sigma^{(0)}+\delta\sigma_{\rm QED}+ \delta\sigma_{\rm rec}+ \delta\sigma_{\rm BW}+ \delta\sigma_{Q}\,,
\end{eqnarray}
where $\sigma^{(0)}$ is the leading-order relativistic result (including the
finite nuclear size effect),
$\delta\sigma_{\rm QED}$ is the QED correction, $\delta\sigma_{\rm rec}$ is the
recoil correction, $\delta\sigma_{\rm BW}$ is the nuclear magnetization distribution
(Bohr-Weisskopf) correction, and $\delta\sigma_Q$ is the electric quadrupole
correction.  

The exact
relativistic result for the leading-order magnetic shielding $\sigma^{(0)}$
was obtained analytically (for a point nucleus) 
\cite{moore:99} and numerically \cite{moskovkin:04}. The recoil
correction is known \cite{rudzinski:09} to the leading order in $\Za$,
\begin{eqnarray}
\delta \sigma_{\rm rec} = -\frac{\alpha \Za}{3}\,\frac{m}{M}\,
\left( 1+ \frac{g_N-1}{g_N}\right)\,,
\end{eqnarray}
where $M$ is the nuclear mass and 
$g_N = M\mu/(\mu_N \,I\,Z\,m_p)$.
The exact relativistic result for the electric-quadrupole correction 
is \cite{moskovkin:04}
\begin{eqnarray}
\delta \sigma_Q = -\frac{\alpha\,(\Za)^3\,Q\,m}{I(2I-1)\,g_I\,m_p}
 \,\frac{6 \left[35+20\gamma-32(\Za)^2\right]}
 {45\,\gamma (1+\gamma)^2\left[15-16(\Za)^2\right]}\,,
\end{eqnarray}
where $Q$ is the nuclear electric quadrupole moment and $\gamma = \sqrt{1-(\Za)^2}$. 

We now turn to the QED correction to the nuclear magnetic
shielding. It consists of the self-energy (SE) and vacuum-polarization 
parts, the SE being the most difficult one. The
Feynman diagrams representing the SE correction (Fig.~\ref{fig:1})
contain two
magnetic interactions, one with the external magnetic field (in what follows,
the Zeeman interaction), 
$V_{\rm zee}(r) = \frac{|e|}{2}\,\bm{B}\cdot (\bfr\times\balpha)$, and the other
with the magnetic dipole nuclear field (in what follows, the hfs interaction), 
$V_{\rm hfs}(r) = \frac{|e|}{4\pi}\, \bm{\mu}\cdot (\bfr\times
\balpha)/r^3$. Formal expressions for the corresponding energy shifts
can be obtained by the two-time Green's function method
\cite{shabaev:02:rep}. Irreducible parts of the diagrams in
Fig.~\ref{fig:1}(a)-(c) give rise to the {\em perturbed orbital} contribution,
\begin{align} \label{po}
\delta E_{\rm po} = 2\,\lbr a |\Sigma(\vare_a) |\delta^{(2)}a\rbr
 + 2\,\lbr \delta^{(1)}_{\rm hfs}a|\Sigma(\vare_a) |\delta^{(1)}_{\rm zee}a\rbr\,,
\end{align}
where $\Sigma$ is the SE operator,  $|\delta^{(1)}_{\rm zee}a\rbr$
and $|\delta^{(1)}_{\rm hfs}a\rbr$ are the first-order perturbations of the
reference-state wave function induced by $V_{\rm zee}$ and $V_{\rm hfs}$,
respectively, and $|\delta^{(2)}a\rbr$ is the second-order perturbation
induced by both interactions. The SE operator is
defined by 
\begin{eqnarray}
\lbr i | \Sigma(\vare)|k\rbr &=& \frac{i}{2\pi}\int_{-\infty}^{\infty} d\omega
  \sum_{n} 
  \frac{\lbr in|I(\omega)|nk\rbr}{\vare-\omega-u\vare_n} \,,
\end{eqnarray}
where 
$I(\omega) = e^2\alpha_{\mu}\alpha_{\nu}D^{\mu\nu}(\omega)$, 
$D^{\mu\nu}(\omega)$ is the photon propagator, and $u \equiv
1-i0$. The diagram in Fig.~\ref{fig:1}(d) gives rise to the {\em hfs-vertex} contribution,
\begin{eqnarray} \label{vrhfs}
\delta E_{\rm vr,hfs} = 2\, \lbr a| \Gamma_{\rm hfs}(\vare_a)  |\delta^{(1)}_{\rm zee} a\rbr
 + 2\,\lbr a| \Sigma^{\prime} |\delta^{(1)}_{\rm zee}a\rbr \,\lbr V_{\rm hfs}\rbr\, ,
\end{eqnarray}
where the prime denotes the derivative of the
operator with respect to the energy argument and 
\begin{align}
\lbr i | \Gamma_{\rm hfs}(\vare)|k\rbr &\ = \frac{i}{2\pi}\int_{-\infty}^{\infty} d\omega
\nonumber \\ & \times
  \sum_{n_1n_2} 
  \frac{\lbr in_2|I(\omega)|n_1k\rbr \lbr n_1|V_{\rm hfs}|n_2\rbr }
 {(\vare-\omega-u\vare_{n_1})(\vare-\omega-u\vare_{n_2})} \,.
\end{align}
The diagram in Fig.~\ref{fig:1}(e) induces the {\em Zeeman-vertex}
contribution, in analogy with its hfs-vertex counterpart,
\begin{eqnarray} \label{vrzee}
\delta E_{\rm vr,zee} = 2\, \lbr a| \Gamma_{\rm zee}  |\delta^{(1)}_{\rm hfs} a\rbr
 + 2\,\lbr a| \Sigma^{\prime} |\delta^{(1)}_{\rm hfs}a\rbr \,\lbr V_{\rm zee}\rbr .
\end{eqnarray}
Finally, Fig.~\ref{fig:1}(f) together with the remaining derivative terms
yields the {\em double-vertex} contribution,
\begin{eqnarray} \label{dvr}
\delta E_{\rm d.vr} &=& 
 2\,\lbr \Lambda\rbr
  + \lbr \Sigma^{\prime\prime}\rbr  \lbr V_{\rm zee}\rbr \lbr V_{\rm hfs}\rbr
 + \lbr \Gamma_{\rm hfs}^{\prime}\rbr \lbr V_{\rm zee}\rbr
\nonumber \\ && 
+ \lbr \Gamma_{\rm zee}^{\prime}\rbr\lbr V_{\rm hfs} \rbr 
+ 2\, \lbr \Sigma^{\prime} \rbr\, \lbr a| V_{\rm zee}|\delta^{(1)}_{\rm hfs}a\rbr,
\end{eqnarray}
where 
$\Lambda \equiv \Lambda(\vare_a)$ is the 4-point vertex operator,
\begin{align} \label{dvra}
\lbr i | \Lambda(\vare)|k\rbr &\ = \frac{i}{2\pi}\int_{-\infty}^{\infty} d\omega
  \sum_{n_1n_2n_3} 
\nonumber \\ & \times
  \frac{\lbr in_3|I(\omega)|n_1k\rbr \lbr n_1|V_{\rm Zee}|n_2\rbr \lbr
    n_2|V_{\rm hfs}|n_3\rbr }
 {(\vare-\omega-u\vare_{n_1})(\vare-\omega-u\vare_{n_2})(\vare-\omega-u\vare_{n_3})}
 \,. 
\end{align}

\begin{table}
\caption{QED corrections to the nuclear magnetic shielding.
 \label{tab:1}}
\begin{ruledtabular}
\begin{tabular}{c..}
 $Z$ &\multicolumn{1}{c}{SE} &
                     \multicolumn{1}{c}{VP}  \\
\hline\\[-7pt]
 10 &   -0.5x1\,(10)   &   0.2x29 \\
 14 &   -0.7x10\,(15)  &   0.2x56 \\
 16 &   -0.7x89\,(9)  &    0.2x71 \\
 20 &   -0.9x27\,(4)  &    0.3x02 \\
 26 &   -1.1x10\,(2)  &    0.3x55 \\
 32 &   -1.2x83\,(1)  &    0.4x17 \\
 40 &   -1.5x19\,(1)  &    0.5x20 \\
 54 &   -2.0x29\,(1)  &    0.7x75 \\
 82 &   -4.4x57\,(2)  &    1.9x96 \\
 92 &   -7.1x07\,(2)  &    2.9x54 \\
\end{tabular}
\end{ruledtabular}
\end{table}

The formulas reported so far refer to the energy
shifts. The corrections to the magnetic shielding 
are related to the energy shifts by 
$\delta \sigma_i = \delta E_i\, IF(F+1)/(\mu\, B\, M_F\, 
\lbr \bm{I}\cdot\bm{F} \rbr)$,
where $M_F$ is the projection of the total momentum $F$. It can be shown that
for the $j = \nicefrac12$ reference states, $\delta \sigma_{\rm QED}$ 
does not depend on nuclear quantum numbers. The numerical calculation of 
$\delta \sigma_{\rm SE}$ was performed along the lines developed in 
Ref.~\cite{yerokhin:08:prl}; its details will be reported elsewhere.

\begin{figure}
\centerline{\includegraphics[width=0.9\columnwidth]{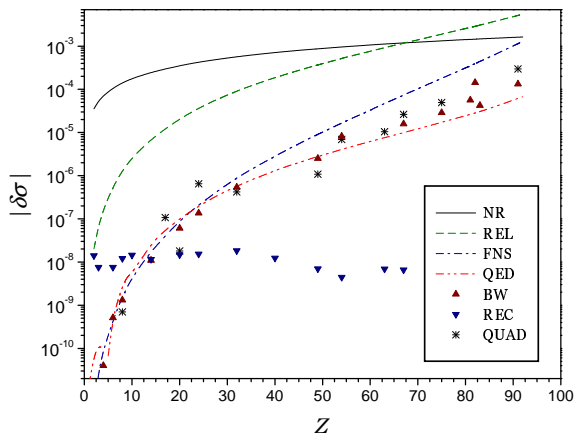}}
\caption{(Color online) Individual contributions to the nuclear shielding.
``NR'' is the nonrelativistic contribution,
  ``REL'' is the relativistic point-nucleus contribution, ``FNS'' is the
  finite nuclear size correction, ``QED'' is the QED correction, ``BW'' is the
  Bohr-Weisskopf correction, ``REC'' is the recoil correction, and ``QUAD'' is the
  electric quadrupole correction. Note that the QED correction changes its
  sign between $Z=4$ and 5.
 \label{fig:2} }
\end{figure}

\begin{table*}
\caption{Individual contributions to the shielding constant $\sigma \times
  10^6$ for selected hydrogen-like ions, see Eq.~(\ref{eq3}).
\label{tab:total}} 
\begin{ruledtabular}
  \begin{tabular}{l.....}
 &  
\multicolumn{1}{c}{$^{17}$O$^{7+}$}        
             &  \multicolumn{1}{c}{$^{43}$Ca$^{19+}$}        
                         &\multicolumn{1}{c}{$^{73}$Ge$^{31+}$}        
                                  &\multicolumn{1}{c}{$^{131}$Xe$^{53+}$}        
                                         &\multicolumn{1}{c}{$^{209}$Bi$^{82+}$}  \\
    \hline\\[-5pt]
Leading        &   143.3x127      &   375.9x60      &   657.x93      &  1461.x6       &  4112x    \\
QED            &    -0.0x026\,(2) &    -0.1x03\,(15)&    -0.x59\,(8) &    -4.x1\,(0.8)&   -30x\,(7) \\
Bohr-Weisskopf &    -0.0x013\,(4) &    -0.0x61\,(18)&    -0.x54\,(16)&    -8.x2\,(2.5)&   -42x\,(13)\\
Quadrupole     &    -0.0x007\,(1) &    -0.0x18      &    -0.x42      &     6.x9\,(0.1)&     7x    \\
Recoil         &    -0.0x120      &    -0.0x15      &    -0.x02      &     0.x0       &     0x    \\
Total          &   143.2x960\,(5) &   375.7x63\,(24)&   656.x36\,(18)&  1456.x3\,(2.6)&  4046x\,(15)\\
  \end{tabular}
\end{ruledtabular}
\end{table*}

The remaining part of the QED effect is the vacuum polarization
(VP). In our calculation, we include two
dominant VP corrections induced by (i) modification of the electron line by the Uehling potential
and (ii) modification of the hfs interaction by the free-loop VP. 

Our calculational results for the SE and VP corrections are
listed in Table~\ref{tab:1}, expressed in terms of the function $D(\Za)$,
\begin{align}     \label{eqDZa}
\delta \sigma_{\rm QED} = \alpha^2\,(\Za)^3\,D(\Za)\,.
\end{align}
The SE correction is calculated for the point
nucleus, whereas the VP part accounts for the finite nuclear size
as well as higher-order iterations of the Uehling potential.
Because of large numerical cancellations, we were able to perform our
numerical SE calculations for $Z\ge 10$ only. In order to
extend our calculations to the lower-$Z$ ions and to cross-check the numerical
procedure, we also performed an analytical calculation of the leading term of
the $\Za$ expansion. The result valid for an $ns$ state reads
\begin{align}    
D_n(\Za) = \frac{8}{9\pi n^3}\, \left[ \ln(\Za)^{-2}
 + 2\,\ln k_{0} -3\,\ln k_{3} -\frac{1817}{480}\right]\,,
\end{align}
where $\ln k_0(1s) = 2.984\,128$ and 
$\ln k_3(1s) = 3.272\,806$ \cite{pachucki:04:prl}. Details of the analytical calculation will be
reported elsewhere. The results of the numerical and the analytical calculations are
in good agreement. 

We now turn to the effect induced by the spatial distribution of the
nuclear magnetic moment, also known as the Bohr-Weisskopf (BW)
correction. Following Ref.~\cite{shabaev:97:pra}, our
treatment of the BW effect is based on the effective
single-particle model of the nuclear magnetic moment. Within this model, the
magnetic moment is assumed to be induced by the odd nucleon with
an effective $g$-factor, which is fitted to yield the
experimental value of the nuclear magnetic moment. 
Under these assumptions, the BW effect can be described by the
magnetization-distribution function $F(r)$ that multiplies the standard point-dipole hfs
interaction $V_{\rm hfs}(r)$. The function $F(r)$ is
induced by the wave function of the odd nucleon, which is obtained by
solving the Schr\"odinger equation with the Woods-Saxon potential
(see Ref.~\cite{yerokhin:08:pra} for details). The BW
correction $\delta \sigma_{\rm BW}$ is obtained by reevaluating the
leading-order magnetic shielding $\sigma^{(0)}$ with the hfs interaction
$V_{\rm hfs}$ multiplied by the magnetization-distribution function $F(r)$. 
The relative uncertainty of 30\% is ascribed to this correction, which is 
consistent with previous error estimates for this effect \cite{shabaev:97:pra}. 

Numerical results of our calculations are presented in Table~\ref{tab:total}
and Fig.~\ref{fig:2}. The error of the QED correction comes from the
numerical uncertainty of the SE part and the estimate of uncalculated VP
terms (30\% of the total VP part). The error of the quadrupole
contribution comes from the nuclear quadrupole moments. The
largest error is due to the BW correction. Since this
effect cannot be presently accurately calculated, this
uncertainty sets the practical limit to which the nuclear magnetic moment can be
determined from an atomic system. For very light
ions, the theoretical accuracy is limited by the recoil
effect (see Fig.~\ref{fig:2}), which is known in the nonrelativistic limit only.
Note that some of corrections to
$\sigma$ depend on the nuclear $g$-factor. This dependence, however, is 
so weak that it can be safely ignored in the determination of the magnetic
moments. 

Summarising, we have presented {\em ab initio} calculations of the
nuclear shielding in hydrogen-like ions, which account for relativistic, 
nuclear, and QED effects. The
present theory permits determination of nuclear magnetic moments with
fractional accuracy ranging from $10^{-9}$ in the case of
$^{17}$O$^{7+}$ to $10^{-5}$ for $^{209}$Bi$^{82+}$. This Letter is
primarily focused on nuclei with spin $I>\nicefrac12$, but the case of 
$I=\nicefrac12$ is only slightly more complicated. Then, the 
nuclear-spin-independent part of $g_F$ in Eq.~(\ref{eq1}) can be cancelled 
approximately, by taking a difference of the $g$-factors $g_F$ for 
two different isotopes of the same element.

Modern experiments on $g$-factors of hydrogen-like ions have 
achieved the accuracy of a few parts in $10^{11}$ \cite{blaum:priv} but so far 
have been restricted to ions with spinless nuclei. Their extention to the nuclei
with spin requires driving the hfs transition and measuring the $g$-factor
of an atom in a hyperfine excited state. These are significant complications but they do
not make an experiment prohibitively difficult \cite{blaum:priv}.

Stimulating discussions with K.~Blaum and G.~Werth are gratefully
acknowledged. Z.H.~was supported by the Alliance Program of the
Helmholtz Association (HA216/EMMI).
V.A.Y.~was supported by the Helmholtz Association
(Nachwuchsgruppe VH-NG-421). K.P.~acknowledges support by NIST 
Precision Measurement Grant PMG 60NANB7D6153.

\end{document}